# Optical properties of excitons in CdSe nanoplatelets

Gerard Czajkowski

Technical University of Bydgoszcz,

Al. Prof. S. Kaliskiego 7, 85-789 Bydgoszcz, Poland

(Dated: October 25, 2025)

**Abstract**

We show how to calculate the linear and nonlinear optical functions of CdSe nanoplatelets, taking into account the effect of a dielectric confinement on excitonic states. We consider both stationary and non-stationary excitation regime. We obtain obtain analytical expressions for the absorption coefficient, the exciton resonance energy and binding energy of nanoplatelets. The impact of plate geometry (thickness, lateral dimension) on the spectrum is discussed. In the nonlinear case we analyze the impact of temperature. For the short-pulse excitation the time dependence of the spectra is considered. The results are compared with the available experimental data.

## 1. INTRODUCTION

The quantum size effects in semiconductor nanocrystals, first pointed out in 1981 [1] have unlocked plethora of research topics for semiconductor nanosystems [2]. The synthesise of nearly monodispersive semiconductor nanocrystallites, such as cadmium selenide (CdSe) [3,4 ] opened the way to producing nanosystems of various dimensions. Here we consider nanoplatelets (NPLs), which are cuboid shaped quantum dots where electrons and holes are confined in three dimensions, of a large lateral size but of only a few molecular layers of thickness. The carriers, electrons and holes, are strongly confined in the growth (z-) direction, and weakly confined in the plane. In the mostly considered CdSe NPL's the vertical confinement is of electrostatic origin and is caused by a large dielectric mismatch between the semiconductor (here CdSe) and its environment. The lateral confinement can be treated similar as in QWs, as a results of impenetrable barrier. Similar as in traditional QDots and QWs, electrons and holes interact by a screened Coulomb potential. The bound electron-hole pairs created by a propagating electromagnetic wave are named excitons and they determine the optical properties of the medium. Cadmium selenide NPLs, first fabricated in 2006 [5] have become important examples of a two-dimensional colloidal nanosystem, with large exciton binding energy, strong quantum confinement and huge oscillator strength, which allows for a high tunability of their optical properties [5]; they exhibit strong and narrow emission lines at both cryogenic and room temperatures. [6] The combined action of very small dimension in the z- direction (few monolayers), strong confinement potential, and Coulomb interaction leads to exciton binding energy reaching hundreds of meV (the bulk binding energy is only 15 meV); it is remarkable quality of CdSe NPLs, which strongly affects their optical properties. Recently several groups have measured the exciton binding energy of CdSe NPLs [6–8]. But the sole knowledge of excitonic binding energy is not sufficient to describe, interpret and explain the observed optical spectra. Beside of the huge binding energy, the are also another interesting effects. For example, Achtstein *et al.* [9] measured the population dynamics of excited state emission from p-states in CdSe NPLs. They also measured the temperature dependence of the emission dynamics. Similar phenomena have been recently analyzed in Cu2O crystals for the even series of Rydberg excitons [10].

The recent growth o interest in such systems encourages us to present a method, which gives a simple analytical expression for optical functions, taking into account confinement and dielectric potentials and any excitonic states,

A majority of authors, describing theoretically the optical properties of excitons in NPLs, are using perturbation calculus, where unperturbed eigenfunctions, describing the carriers motion in the z-direction, are the (finite or infinite) 1-dimensional quantum well functions. The binding energy is then calculated with Coulomb e-h interaction potential considered as perturbation [11,12]. A numerical attempt to calculate electronic properties of CdSe NPLs spectra has also been taken by Benchamekh,*et* al. [6] who applied an advanced tight-binding model. In the present work we propose a different confinement potential, resulting directly from the dielectric confinement. This potential allows for an analytical solution of the Schrödinger equation for electron and hole, describing their motion in the z-(confinement) direction. We obtain both eigenfunctions and eigenvalues. Adding the solution of the Schrödinger equation for in plane relative motion, we can estimate the total binding energy. Moreover, we present the theoretical method, which allows one to calculate optical properties (i.e., positions of resonances and absorption spectra) of NPLs depending on numbers of monolayers (i.e. thickness of NPLs) and the lateral extension. In calculations we have used the so-called real density matrix approach (RDMA), which allows one to obtain analytical expressions for the susceptibility. The method has been used extensively in bulk semiconductor materials, and has been also applied to various types of nanostructures, to compute linear and nonlinear optical properties (see, for example, Refs.13–15).

The first attempt to apply RDMA in the case of CdSe NPLs is given in Ref. [16]. We extend the approach from Ref.16, adding results on the exciton states size dependence, and nonlinear effects, such as exciton population decay, and temperature dependence of the spectra.

The paper is organized as follows. In Sec. 2 we recall the basic equations of the used approach (RDMA). In Sec. 3 we present an approximation, which enables analytic calculations. The results obtained by the method are presented in Sections 4 and 5,, containing excitonic resonance energies, binding energies, and absorption spectra for NPLs with different sizes, including the time- and temperature dependence of the absorption spectra. The last section presents the concluding remarks. Appendix A contains the details of analytical calculations.

## 2. THEORY

We consider a CdSe nanoplatelet of cuboid shape, located at the z = 0 plane, and with the barriers located at x $=\pm L_x/2$; y$=\pm L_y/2$; z$=\pm L_z/2$. We consider the response of the NPL to a normally incident electromagnetic wave, linearly polarized in the x-direction

$$\mathbf{E}_i(z; t) = \; = \mathbf{E}_{i0}(t) \exp(ik_0 z - i\omega t); \quad k_0 = \omega/c. \tag{1}$$

Since we will consider both stationary, and nonstationary excitation, the amplitude $\mathbf{E}_i(t)$ is assumed in the form

$$\mathbf{E}_i(z,t) = \mathbf{E}_{i0} \, F(t) \exp(ik_0 z - i\omega t); \tag{2}$$

where F(t) describes the pulse shape in the case of nonstationary excitation. For a stationary excitation we put F(t) = 1. The calculations of the optical response in the framework of RDMA is based on the solution of the so-called constitutive equations for two-point correlation functions $Y(\mathbf{x_e;x_h})$ (interband transition amplitude, also called exciton amplitude), and $C(\mathbf{x_e;x_h})$, $D(\mathbf{x_e;x_h})$ (intraband transitions), where $\mathbf{x_e}$; $\mathbf{x_h}$ are the electron and hole coordinates. The derivation and the explicit form of the constitutive equations can be found in Ref.[13]. The constitutive equations must be solved simultaneously with the Maxwell field equations

$$-c^2\varepsilon_0 \nabla \times \nabla \times \mathbf{E} - \varepsilon_0\varepsilon_b \partial_t^2 \mathbf{E} = \partial_t^2 \mathbf{P}, \qquad (3)$$

where $\mathbf{P}$ ist he polarization, and $\varepsilon_b$ the bulk dielectric constant. When the effects of confinement are considered, one makes use of the appropriate boundary conditions on $\mathbf{E}$, Y, C, and D.

In the weak excitation field limit and for the linear case, where we put C = D = 0 on the right-hand side, we obtain equations for the intraband transition amplitudes $Y_{12}^{\alpha\,b}$ of the electron-hole pair of coordinates $x_1 = x_h$ and $x_2 = x_e$ between any pair of bands $\alpha$ and b. The equations have the form have the form

$$-i(\hbar\partial_t + \Gamma_{\alpha\,b})Y_{12}^{\alpha\,b} + H_{eh\alpha b}\, Y_{12}^{\alpha\,b} = \mathbf{M}_{\alpha\,b}\, \mathbf{E}, \qquad (4)$$

where $\Gamma_{\alpha b} = \hbar/\tau_{\alpha b}$, $\tau_{\alpha b}$ is the exciton life time [13], and $\hbar = 0.6582$ meV $\times$ ps is the Planck's constant.

The two-band Hamiltonian $H_{eh\alpha b}$ with the energy gap $E_{g\alpha b}$ for any pair of bands reads

$$H_{eh\alpha b} = E_{g\alpha b} + (1/2m_\alpha)\,\mathbf{p}_{h\alpha}^2 + (1/2m_b)\,\mathbf{p}_{eb}^2 + V_{eh}(1,2) + V_h(1) + V_e(2), \qquad (5)$$

With the electron and hole kinetic energy operators, $m_\alpha$ and $m_b$ being the band effective masses, $V_{eh}$ describes the electron-hole attraction and $V_e, V_h$ denote the confinement potentials of the electron and the hole, respectively. The total polarization of the medium is related to the excitonic amplitudes by

$$\mathbf{P(X)} = \sum_{\alpha,\dots,a,\dots}\mathrm{Re}\int d^3x\,\mathbf{M}_{\alpha\,b}\,Y_{12}^{\alpha\,b}(\mathbf{X,x}), \qquad (6)$$

where $\mathbf{x=x_e-x_h}$ is the relative electron-hole coordinate, X is the electron-hole pair center-of-mass coordinate, and the summation includes all allowed excitonic transitions between the valence and conduction bands.

For CdSe based NPLs we have to consider both heavy(H) and light(L) hole excitons. For the optical transitions between (a=H,L_) valence bands and the conduction band (b=C) we get two constitutive equations for the excitonic amplitudes $Y_{12}^{\alpha\,C} = Y_\alpha(\mathbf{x_e,x_h})$

$$-i\,\hbar\,\partial_t\,Y_\alpha - i\Gamma_\alpha + H_{eh\alpha}\,Y_\alpha = \mathbf{M}_\alpha(\mathbf{x})\,\mathbf{E(X)}, \qquad (7)$$

where $\mathbf{M}_\alpha(\mathbf{x})$ are transition dipole densities. Considering CdSe based NPLs we have to account the effective masses anisotropy. In consequence, the operators $H_{eh\alpha}$ have the form

$$H_{eh\alpha} = E_{g\alpha} + p_{hz}^2/2m_{hz\alpha} + V(1) + p_{ez}^2/2m_{ez} + V(2) + \mathbf{P}_X^2/2M_{\|\alpha} + \mathbf{P}_Y^2/2M_{\|\alpha} + \mathbf{p}_{\rho e}^2/\mu_{\|\alpha}$$
$$+ \mathbf{p}_{\rho h}^2/\mu_{\|\alpha} + V(z_e - z_h, \boldsymbol{\rho}), \qquad (8)$$

where V($z_e-z_h$, **ρ**) is the screened Coulomb interaction

$$V(z_e-z_h, \boldsymbol{\rho}) = -e^2\{4\pi\varepsilon_0\varepsilon_1[\rho^2 + (z_e-z_h)^2]^{1/2}\}^{-1} , \qquad (9)$$

with $\varepsilon_1$ being the bulk CdSe delectric constant, and $\boldsymbol{\rho}_{e,h}$ are two-dimensional vectors in the x-y plane,

$$\boldsymbol{\rho}_{e,h} = (x_{e,h}, y_{e,h}), \qquad \rho^2 = (\boldsymbol{\rho}_e - \boldsymbol{\rho}_h)^2 = (x_e - x_h)^2 + (y_e - y_h)^2. \qquad (10)$$

We have separated the center of mass coordinates $\mathbf{X}_\parallel$, $\mathbf{Y}_\parallel$ and the related momenta $P_X$, $P_Y$ from the relative coordinates $\boldsymbol{\rho}_e$, $\boldsymbol{\rho}_h$ on the (x,y) plane and the related momenta, $M_{\parallel\alpha} = m_{h\parallel\alpha} + m_{e\parallel}$ is the total in plane excitonic mass, $\boldsymbol{\rho}_e - \boldsymbol{\rho}_h$ is the relative coordinate in the x–y plane. Using Hamiltonian (8), with the harmonic time dependence $\sim \exp(-i\omega t)$, we obtain the constitutive equations in the form

$$(H_{eh\alpha} - \hbar\omega - i\Gamma_\alpha) Y_\alpha(\boldsymbol{\rho}_e, \boldsymbol{\rho}_h, z_e, z_h) = \mathbf{M}_\alpha(\boldsymbol{\rho}, z_e, z_h)\mathbf{E}(\mathbf{X}_\parallel, Z), \qquad (11)$$

Where $\hbar\omega$ is the energy of the exciting wave. Note that the above equation refers to a 6-dimensional space. In spite of the symmetry differences between the onfinement potentials and the Coulomb potential, an analytic solution of Eq. (11) not exists. Therefore several approximations have been used. One of them is to seek the solution as a product of eigenfunctions of one dimensional infinite-deep confinement potential

$$\psi(x_i^{c,v}) = (2/L_i)\sin(\pi n_i^{c,v}/L_i). \qquad (12)$$

The functions (12) are then used within the framework of perturbation theory [11,12].

In the next section we show how, with certain simplifying assumptions, the dimensionality of the problem can be reduced from 6 to 3, enabling to obtain analytical solution of the linear inter-band equation (11) with respective exciton resonance energies.

## 3. Approximate solution of the linear constitutive equation

To reduce the dimensionality of our problem, following simplifications are assumed.
1. Since, in the case of CdSe, the effective mass of the hole is much larger than that of the electron, we neglect the effects of the lateral motion of the hole, considered it as located in the center of the NPL. Similar assumptions were made in the past in the case of Quantum Dots [17], and Quantum Disks [16,18].
2. We assume, that the optical properties of a rectangular NPL can be described by a motion of the electron in a cylindrical disk with the radius

$$r_{eff} = (L_x L_y/\pi)^{1/2} . \qquad (13)$$

3. The confinement potential for the motion of the electron in x - y plane is given by expression

$$V_e(\rho_e) = 0 \quad \text{for} \quad \rho_e \leq R,$$
$$= \infty \quad \text{for} \quad \rho_e > R, \qquad (14)$$

where we put $R = r_{eff}$. For the further calculations we will need the eigen-functions and eigenvalues of a two-dimensional Schrödinger equation

$$[\mathbf{p}^2_\| /2m_{e\|} - e^2/4\pi\varepsilon_0\varepsilon_1\rho_e + V_e(\rho_e)]\ \psi(\rho_e) = E\ \psi(\rho_e)\ .\qquad(15)$$

satisfying the boundary condition $\psi(\rho_e) = 0$ for $\rho_e = R$. The electron inside the NPL is subject to two competing forces: the repulsive (confinement) force, connected with a positive energy, and the attractive Coulomb force (attraction from the hole), related to a negative energy. Therefore the eigenenergy $E_{jm}$, being the sum of the mentioned contributions, can be positive, negative, or zero. For the negative energies, we have [18]

$$\psi_{jm}(\xi,\phi) = C\ \xi^{|m|}\ e^{-\xi/2}\ \times M(m+1/2-\eta;2|m|+1;\ \xi)\ e^{im\phi}/(2\pi)^{1/2}\ ,\qquad(16)$$

$j$ and $m$ are the principal and magnetic quantum numbers of the excitonic state, $C$ is the normalization constant, $\rho_e/a_{e\|}{}^* = \rho$,

$$\eta=2/k,\qquad \xi=k\rho,\qquad k^2 = -4\ (2\ m_{e\|}/\hbar^2)\ a_{e\|}{}^{*2}E,\quad a_{e\|}{}^* = (m_0/m_{e\|})\ \varepsilon_1 a_B{}^*,$$

where $m_0$ is the free electron mass, and $a_B{}^* = 0.0529$ nm the hydrogen Bohr radius. The quantities $k$, $\rho$, and $\xi$ are dimension-less. The eigenfunction, due to the boundary condition (14), satisfies the equation

$$\psi_{jm}(k\rho_{eff},\phi) = 0,\qquad \rho_{eff} = R/\ a_{e\|}{}^*,\qquad(17)$$

giving the eigenenergies $E_{jm}$, $j=0,1,2,\ldots$, $m=0,1,2,\ldots$ In the case of positive eigenenergy, the eigenfunction has the form

$$\psi_{jm}(\xi,\phi) = C\ \xi^{|m|}\ e^{-i\xi/2}\ \times M(m+1/2 - i\eta;2|m|+1;\ i\xi)\ e^{im\phi}/(2\pi)^{1/2}\ ,\qquad(18)$$

with the normalization constant $C$. The eigenenergy can be calculated by the condition

$$\text{Re}\ [\ \psi_{jm}(\rho_{eff})]=0.\qquad(19)$$

The positive eigenenergies are then given by

$$E_{jm} = (1/\eta^2)\ R^*_{e\|}\ ,\qquad\qquad R^*_{e\|} = (m_{e\|}/m_0)(1/\varepsilon_1{}^2)R^*_B,\qquad(20)$$

$R^*_B = 13600$ meV being the hydrogen Rydberg energy. In the above equations $M(a;b;z)$ is the confluent hypergeometric function (notation by Ref. [19]).

4. The movement of electrons and the holes in the z-direction is affected by the dielectric confinement potential, which we take in the form

$$V_{e,h}(z) = \gamma_{e,h}[(L_z/2) - z]^{-1}\ ,\qquad(21).$$

The coefficient $\gamma$ is proportional to dielectric coefficients

$$\gamma = (\varepsilon_1 - \varepsilon_2)[\varepsilon_1(\varepsilon_1 - \varepsilon_2)]^{-1},$$

where $\varepsilon_2$ and $\varepsilon_2$ are the dielectric constants of external and internal media, respectively, and $\varepsilon_2 \ll \varepsilon_1$. Using the potential (21), we solve the Schrödinger equation

$$- (\hbar^2/2m_z)\, d_z^2 \psi_{e,h}(z) \; + \gamma\, R^*_z\, a^*_z[(L/2) - z]^{-1}\psi_{e,h}(z) = E_z \psi_{e,h}(z), \qquad (22)$$

where $m_z$ denotes the effective mass of the considered quasi particle (electron or hole) in the z-direction, $R^*_z, \le a^*_z$ are the corresponding effective Rydberg energies and Bohr radii defined as

$$R^*_{e,hz} = (m_{e,hz}/m_0)\,(1/\varepsilon_1^2)R^*_B, \qquad (23)$$

$$a_{e|,hz}^* = (m_0/m_{e\|})\,\varepsilon_1\, a_B^*. \qquad (24)$$

The eigenfunctions resulting from (22) have the form [18]

$$\psi\,(u) = C\, u\, e^{-u/2}\, M(1-\lambda;2;u), \qquad (25)$$

with the normalization C. The following notation is used

$$d = L_z/2a^*_z, \quad \zeta = (d-z)/a^*_z, \quad \lambda = -\, i\gamma/k,$$
$$\varepsilon = E/R^*, \quad k = 2\sqrt{\varepsilon}, \quad u = -i\, k\, \zeta. \qquad (26)$$

The eigenvalues follow from the condition $\psi(-ikd) = 0$ and, in the lowest appro-ximation, have the form

$$E_{e,hz} = 6\,\beta[1+(\gamma_{e,h}\, d_{e,h}/2)]/(m_{e,hz}\, L_z^2), \qquad (27)$$

with

$$\beta = a^{*2}_B \times R^*_B = 38\ \text{nm}^2\,\text{meV}, \qquad (28)$$

and $L_z$ is expressed in nm. Using the above defined eigenfunctions, we define $Y(\rho_e, z_e, z_h)$ as a series

$$Y = \sum_{jm} Y_{0jm}\, \psi_{jm}(\rho_e)\, \psi_e(z_e)\, \psi_h(z_h), \qquad (29)$$

with certain coefficients $Y_{0jm}$. We restricted the consideration to the lowest confinement functions in the z direction. Substituting the expansion (29) into Eq. (11) we obtain the equation

$$HY = \mathbf{ME} - (\Delta\, V)Y, \qquad (30)$$

where

$$H = E_g + p^2_{ez}/2m_{ez} + V_e(z_e) \; - i\Gamma$$

$$+ \quad V_h(z_h) + \mathbf{p}^2_{\parallel}/2m_{e\parallel} + V_e(\rho_e) \ -e^2/4\pi\varepsilon_0\varepsilon_1\rho_e \tag{31}$$

and

$$\Delta V = e^2/4\pi\varepsilon_0\varepsilon_1\rho_e \ - e^2\{4\pi\varepsilon_0\varepsilon_1[\rho_e{}^2 + (z_e-z_h)^2]^{1/2}\}^{-1} \quad , \tag{32}$$

where $E_g$ ist the fundamental gap energy.

Eq. (30) can be solved by means of a Green's function G appropriate to the l.h.s operator in (30), satisfying the equation

$$[\ E_g \ - \ \hbar\omega - i\Gamma + p^2_{ez}/2m_{ez} + V_e(z_e) \quad + \ p^2_{hz}/2m_{hz} + V_h(z_h) + \mathbf{p}^2_{\parallel}/2m_{e\parallel} + V_e(\rho_e)$$

$$-e^2/4\pi\varepsilon_0\varepsilon_1\rho]\ G(\rho,\sigma;\ \phi,\varphi;z_e,w_e,z_h,\ w_h) \quad = \ - \ (1/2\pi\rho)\ \delta(\rho-\sigma)\ \delta(z_e-w_e)\ \delta(z_e-w_h)\ \delta(\phi-\varphi). \tag{33}$$

**Table 1**: Masses, reduced masses, Rydberg energies, Luttinger parameters, and coherence radii, from Ref. [16].

| Parameter | 3ML | 4ML | 5ML |
|---|---|---|---|
| L | 1 | 1.33 | 1.67 |
| $m_{ez}$ | 0.2567 | 0.2015 | 0.1635 |
| $m_{e\parallel}$ | 0.3208 | 0.2519 | 0.2044 |
| $m_{hzH}$ | 1.1925 | 0.9754 | 0.8153 |
| $m_{h\parallel H}$ | 0.4957 | 0.4337 | 0.3879 |
| $m_{hzL}$ | 0.4149 | 0.3659 | 0.3302 |
| $m_{h\parallel L}$ | 0.8121 | 0.6887 | 0.5963 |
| $\mu_{zH}$ | 0.2112 | 0.1670 | 0.1362 |
| $\mu_{\parallel H}$ | 0.1948 | 0.1593 | 0.1338 |
| $\mu_{zL}$ | 0.1586 | 0.1300 | 0.1094 |
| $\mu_{\parallel L}$ | 0.2300 | 0.1844 | 0.1522 |
| $R^*_{\parallel H}$ | 73.58 | 60.20 | 51.28 |
| $R^*_{\parallel L}$ | 86.88 | 69.67 | 57.49 |
| $\gamma_1$ | 1.6243 | 1.8789 | 2.1062 |
| $\gamma_2$ | 0.3929 | 0.4269 | 0.4488 |
| $\rho_{0H}$ | 0.20 | 0.18 | 0.17 |
| $\rho_{0L}$ | 0.22 | 0.19 | 0.18 |

The Green function can be expressed in terms of the eigenfunctions of the operator (31), and has the form

$$G = (1/2\pi)\ )\ \psi^*_e(z_e)\ \psi_h(w_e)\ \psi^*_h(z_h)\ \psi_h(w_h) \times \textstyle\sum_{jm} (1/\ k_{jm}{}^2)\ \psi^*_{jm}(k\rho)\ \psi_{jm}(k\sigma), \tag{34}$$

where

$$k_{jm}{}^2 = (2m_{e\parallel}/\ \hbar^2)\ (E_g - \hbar\omega - i\Gamma \ +E_{conf} + E_{jm}). \tag{35}$$

The confinement energies $E_{conf}$ result from Eq. (27)

$$E_{conf} = E_{ez} + E_{hz}. \tag{36}$$

The excitonic amplitude, obtained from Eq. (30) by means of Green's function (34), has the form

$$Y = G \, \mathbf{ME} - G \, (\Delta V)Y. \tag{37}$$

From the above equation one obtains the coefficients $Y_{0jm}$,

$$Y_{0jm} = (1/Q_{jm}) \quad < \psi_e(z_e) \, \psi_h(z_h) \, \psi_{jm}(k\rho)| \, \mathbf{ME} > \quad . \tag{38}$$

The resonant denominators $Q_{jm}$ are given as

$$Q_{jm} = E_g + E_{conf} + E_{jm} + <2/\rho> \quad - <2/[\rho^2 + (z_e - z_h)^2]^{1/2} > - \hbar\omega - i\Gamma. \tag{39}$$

The expressions $<2/\rho>$ and $<2/[\rho^2 + (z_e - z_h)^2]^{1/2}>$ are defined as follows

$$<2/\rho> = 2R^*_{e\parallel} \int_{NPL} \psi^2_{jm}(\rho) \, \psi^2_e(z_e) \, \psi^2_h(z_h) d\rho \, dz_e \, dz_h, \tag{40}$$

$$E_{bjm} = -<2[\rho^2 + (z_e - z_h)^2]^{-1/2} > = -2 \, R^*_{e\parallel} \int_{NPL} \psi^2_{jm}(\rho) \, \rho d\rho \tag{41}$$

$$\times \psi^2_e(z_e)\psi^2_h(z_h)dz_e dz_h \, [\rho^2 + (z_e - z_h)^2]^{-1/2},$$

where $E_{bjm}$ is called the exciton binding energy. The limits of integration include the volume of NPL, i.e. $0 \le \rho \le \rho_{eff} - L_z/2 \le z_e, z_h \le L_z/2$. Introducing the total confinement energy

$$E_{conf,tot,jm} = E_g + E_{conf} + <2/\rho> \quad + E_{jm}, \tag{42}$$

we rewrite Eq. (39) to the form

$$Q_{jm} = E_{res,jm} - \hbar\omega - i\Gamma, \tag{43}$$

where

$$E_{res,jm} = E_{conf,tot,jm} + E_{bjm} \tag{44}$$

is the exciton resonance energy at the state jm. All the calculations can be performed for heavy- and light hole excitons, with using the appropriate parameters.

**TABLE 2**: Binding energies and exciton resonance energies calculated for lateral 5ML disk, thicknes $L_z = 1{:}67$ nm, analyzed by Achtstein. et al. [9], lengths in nm, masses in free electron mass, energies in meV, the energy gap at 4 K temperature 1840 meV, notation: 1: 8:1×3:6, 2: 17×6, 3: 21×7, 4:29×8, 5: 41×13, 6: 30×15.

| Lat.extension | 1 | 2 | 3 | 4 | 5 | 6 |
|---|---|---|---|---|---|---|
| $m_{e\parallel}$ | 0.2044 | 0.2044 | 0.2044 | 0.2044 | 0.2044 | 0.2044 |
| $R^*_{e\parallel}$ | 77.2 | | | | | |
| $a^*_{e\parallel}$ | 1.553 | | | | | |
| $r_{eff}$ | 3.05 | 5.7 | 6.84 | 8.59 | 13.13 | 11.96 |
| $\rho_{eff}$ | 1.96 | 3.67 | 4.4 | 5.53 | 8.38 | 7.7 |
| $E_{conf,tot}(1SH)$ | 1097.68 | 1013 | 1001 | 993 | 986.5 | 979.62 |
| $E_{conf,tot}(2SH)$ | 1210.18 | 793.77 | 764.27 | 741.32 | 723.04 | 720.11 |
| $E_{conf,tot}(1PH)$ | 839.88 | 760.84 | 744.4 | 729.7 | 716.55 | 714.4 |
| $E_{conf,tot}(2PH)$ | 831.2 | 769.94 | 768.47 | 768.57 | 727.29 | 722.46 |
| $E_{conf,tot}(1SL)$ | 1249.68 | 1165 | 1153 | 1145 | 1138.5 | 1131.62 |
| $E_{conf,tot}(2SL)$ | 1362.18 | 945.77 | 916.27 | 893.32 | 875 | 872.11 |
| $E_{conf,tot}(1PL)$ | 991.8 | 912.84 | 896.4 | 881.7 | 868.55 | 866.4 |
| $E_{conf,tot}(2PL)$ | 1346.2 | 921.94 | 920.47 | 920.57 | 879.29 | 874.46 |
| $|E_b(1SH)|$ | 391.16 | 362 | 359.3 | 358 | 357.3 | 354.88 |
| $|E_b(2SH)|$ | 299.4 | 104 | 87.3 | 75.4 | 67.7 | 66 |
| $|E_b(1PH)|$ | 153.9 | 100.8 | 91.5 | 83 | 75.14 | 74 |
| $|E_b(2PH)|$ | 100 | 104.7 | 102.9 | 101.14 | 69.5 | 59.91 |
| $E_{res}(1SH)$ | 2555.11 | 2491 | 2481.7 | 2475 | 2469.2 | 2464.7 |
| $E_{res}(2SH)$ | 2750.8 | 2529.7 | 2517 | 2506 | 2495.3 | 2494.1 |
| $E_{res}(1PH)$ | 2493 | 2500 | 2493 | 2486.7 | 2480.4 | 2481.6 |
| $E_{res}(2PH)$ | 2571.2 | 2503.0 | 2507 | 2505.6 | 2503.53 | 2492.9 |
| $E_{res}(1SL)$ | 2707.11 | 2643 | 2633.7 | 2627 | 2621.2 | 2616.7 |
| $E_{res}(2SL)$ | 2902.8 | 2681.7 | 2669 | 2658 | 2647.3 | 2646.2 |
| $E_{res}(1PL)$ | 2645 | 2652 | 2645 | 2638.7 | 2633.5 | 2632.3 |
| $E_{res}(2PL)$ | 2723.2 | 2659.4 | 2659.25 | 2659.4 | 2656.6 | 2655.55 |

With the use of the above quantities we obtain the amplitudes Y, determining the NPL polarization by Eq. (6), from which the mean NPL susceptibility is calculated, having the form

$$\chi = (2/\varepsilon_0) \sum | < M(r, z_e, z_h) |\psi_{jm}(\rho)\psi_e(z_e)\psi_h(z_h)>|^2 \times (E_{res} - \hbar\,\omega - i\Gamma)^{-1} \quad . \qquad (45)$$

The susceptibility (45) enables one to obtain the NPL's optical functions (absorption, reflection, and transmission coefficients).

## 4. Results of specific calculations. Stationary excitation

We have performed calculations for CdSe NPLs having in mind the experimental results by Brumberg *et al*. [12], and Achtstein *et al*. [9] . In this works two types of measurements are performed. In Ref. [9] PL emission of CdSe NPLs with fixed thickness ($L_z$) and various lateral dimensions has been measured. In Ref. [12] absorption spectroscopy in pulsed magnetic fields for three different CdSe NPL thicknesses and lateral areas is implemented.

Here we discuss the case without magnetic field. We analyze the both cases, calculating the size dependent exciton states energies, and displaying the related emission/absorption line shapes. The basic ingredients used in the calculations are the band-gap data (band-gap energy, electron and hole effective masses, longitudinal-transverse energy), supplemented with the 'in' and 'out' dielectric constants $\varepsilon_1$ and $\varepsilon_2$. We follow the estimation of electron and hole effective masses, given in Ref. [16]. Other parameters are collected in Tables 1–5.

We start with the results from Ref. [9], where the NPLs of thickness 4.5 ML (we take 5 ML in calculations) and lateral dimensions ranging from 8 × 3:6 to 30×15, are considered, (dimensions in nm). Using Equations (41) and (44) we have computed the exciton resonance energies and binding energies of heavy-hole exciton states 1SH,L (j = 0; m = 0) , 2SH,L (j = 1; m = 0), and 1PH,L (j = 0; m = 1, 2PH,L (j=1, m=1) appropriate to the heavy- and light hole excitons (α = H,L). The results are presented in Table S1. Comparing the theoretical and experimental results from Ref. [9]  (Table S2) we observe a very good agreement. In the next step we calculate the NPLs absorption spectrum.

In the next step we calculate the NPLs absorption and emission spectra. In the considered NPLs widths the typical wavelength of the input electromagnetic wave is much larger than the NPLs width, so we can use the long wave appro-ximation. For further calculations we need to define the dipole density function **M**. The transition dipole density **M(r)** should
have the same symmetry properties as the solution of the corresponding Schrödinger equation. Since we focus our attention on S and P states, we assume, the dipole densities  in the form

$$\mathbf{M}_{GS}(\mathbf{r}) = \mathbf{M}_{0;GS} N_{GS}\, e^{-\rho/\rho 0} \delta(z_e - z_h) \qquad (46)$$

for S states (GS =ground state),  and

$$\mathbf{M}_{ES}(\mathbf{r})\ = \mathbf{M}_{0;ES} N_{ES}\ \times \rho\, \exp(\text{-}\rho/\rho_0) \delta(z_e - z_h)\, (2\pi)^{-1/2}\, e^{im\phi}, \qquad (47)$$

for the P state (ES=excited state),  where $\rho_0$ are the so-called coherence radii,

$$\rho_{0H}=(R^{*}_{\|H}/\, E_g)^{1/2}, \qquad\qquad \rho_{0L}=(R^{*}_{\|L}/\, E_g)^{1/2}, \qquad (48)$$

and $N_{GS}$ ,$N_{ES}$ are normalization constants such that, for example

$$N_{SH} \int \rho \exp\, (\, -\rho/\rho_{0H})\, d\rho = 1, \qquad (49)$$

integrated in the limits (0, $\rho_{eff}$).

The dipole matrix elements $\mathbf{M}_{0;GS}$; $\mathbf{M}_{0;ES}$ are known in the case of bulk semiconductors, and their values are related to the so-called Longitudinal-Transversal splitting energy. The latter can be measured using the polariton dispersion relation. In the quasi 2-dimensional systems with lateral confinement, and for normal incidence, we have no polaritons, and the dipole matrix elements are unknown. To estimate them, we use an expression based on the definition of the matrix elements between the hole and electron states of all allowed transitions $j_1$, $m_{1,}$ $j_2$ , $m_2$ ($p^e_x$, $p^h_x$, $s^e$, $s^h$), see Ref. [9],

$$\mathbf{M}_{0j1m1,j2m2}\ = -e \int d^2\rho\ \psi_{j1,m1}\, (\rho)\ \boldsymbol{\rho}\ \psi_{j2,m2}\, (\rho), \qquad (50)$$

where $e$ is the electron charge. The results for the ground and excited states, and various NPL extensions, are given in Tables 2-5.

**TABLE 3**. Sizes and exciton states energies of CdSe NPLs from Ref. [9].

| Lat. Extens. | 1SH | 2SH | 1PH | 2PH | 1SL |
|---|---|---|---|---|---|
| 8:1 × 3:6 | 2574.6 | 2750 | 2512 | | |
| 17 × 6 | 2491 | 2530 | 2500 | 2505 | 2643 |
| 21 × 7 | 2482 | 2517 | 2493 | 2505 | 2634 |
| 29 × 8 | 2475 | 2506 | 2487 | 2507 | 2627 |
| 30 ×15 | 2469 | 2495 | 2481 | 2498 | 2621 |
| 41 ×13 | 2464.7 | 2494 | 2480 | 2502 | 2616.7 |

**TABLE 4**: Exciton resonance energies at room temperature, calculated for lateral 5ML disk with the energy gap at room temperature (1750 meV), notation: 1: 8:1 ×3:6, 2: 17 × 6, 3: 21 × 7, 4: 29 × 8, 5: 41 × 13, 6: 30 × 15.

| Lat.exts. | 1 | 2 | 3 | 4 | 5 | 6 |
|---|---|---|---|---|---|---|
| $E_{res}$(1SH) | 2484.6 | 2401 | 2391.7 | 2385 | 2379.2 | 2374.7 |
| $E_{res}$(2SH) | 2660.8 | 2439.7 | 2427 | 2416 | 2405.3 | 2404.1 |
| $E_{res}$(1PH) | 2435.9 | 2410 | 2403 | 2396.7 | 2390.4 | 2391.6 |
| $E_{res}$(2PH) | 2416 | 2413 | 2417 | 2417 | 2408 | 2412 |
| $E_{res}$(1SL) | 2636.6 | 2553 | 2543.7 | 2537 | 2531.2 | 2526.7 |
| $E_{res}$(2SL) | 2812.8 | 2591.7 | 2579 | 2568 | 2557.3 | 2556.2 |
| $E_{res}$(1PL) | 2587.9 | 2562 | 2555 | 2548.7 | 2548.7 | 2542.3 |
| $E_{res}$(2PL) | 2568 | 2569.4 | 2569.25 | 2569.4 | 2566.6 | 2565.55 |

Having determined the dipole transition densities, we calculate the oscillator strengths (the radial part) related to given transition, by the relation

$$f_{jm} = \left| \int \rho \, d\rho \, M(\rho) \, |\psi_{jm}(\rho) \, |^2. \right. \tag{51}$$

The radial damping parameter, displayed in Table 5, result from fitting to results of Ref. [9], where one reads $\Gamma_{ES,r} = 10.32$ meV for $29× 8$ extension, and $32.24$ meV for $41 × 13$ nm. The ground state damping parameters are 2.43 meV and 4.2 meV for analogous areas. The above values can be fitted by expressions

$$\Gamma_{ES} = A_{ES} \exp(-0.0008 \rho_{eff}^2 + 0.2692 \, \rho_{eff}), \quad \Gamma_{GS} = A_{GS} \exp(0.0084 \rho_{eff}^2 - 0.0628 \, \rho_{eff}),$$

$$A_{ES} = 1.102, \quad A_{GS} = 2.2567. \tag{52}$$

With the above definitions we have all elements to calculate the mean NPL susceptibility, which can be written in the form

$$\chi = (2/\varepsilon_0) \sum_{jm} f_{jm} \, |M_{0,jm}|^2 (E_{res,jm} - \hbar \, \omega - i\Gamma)^{-1}, \tag{53}$$

with the summation including considered transitions. The experiments of Ref. [9] were performed for excitation energy in the region of the 1SH (Ground state GS) and 1PH (Excited state ES) excitonic resonances. Therefore, using Equation (53), we calculate the absorption coefficient as the imaginary part of the susceptibility, taking into account the above mentioned states. With regard to normalization, we obtain the absorption coefficient by the formula

$$\alpha = \Gamma^2_{GS}[(E_{res,GS}-E_{in})^2 + \Gamma^2_{GS}]^{-1} + K_s\, \Gamma_{GS}\, \Gamma_{ES}\, [(E_{res,ES}-E_{in})^2 + \Gamma^2_{ES}]^{-1}, \quad (54)$$

where s counts the NPLs (s=1,…,6, see Table 2)

$$K_s = f_{ES}\, M_{0,ES}^2\, /f_{GS}\, M_{0,GS}^2, \quad (55)$$

and $E_{in} = \hbar\omega$ ist the excitation energy. The results are displayed in Figures 1-6. In Figure 1 we observe an inversion of states: the excited state is red shifted compared to the ground state. This is an effect which can be attributed to the competition of the confinement and the Coulomb attraction. For the smallest NPL the radius $r_{eff}$ is smaller than the critical radius for the P excitons (see Appendix), it means that the confinement energy prevails over the Coulomb energy, which in effect gives the shift of resonance towards lower energies. For larger NPL extensions the situation is 'normal', it means the P exciton (ES) resonance has a larger energy than the GS resonance. We observe a fairly good agreement with experimental results by Achtstein *et al.* [9]. We also observe a decreasing of ES-GS intensity ratios with increasing NPL area, as was observed in experiments.

When the excitation energy interval increases, higher excited states should be taken into account. In particular, we observe the appearance light-hole 1SL and 1PL excitonic resonances, which have comparable oscillator strengths with the 1SH and 1PH excitons. It is illustrated in Figures 7-11.

**TABLE 5**: Oscillator strengths and damping parameters for 1S, 2S, and 1P excitons for NPLs with dimensions as in Table 1: 1: 8:1 ×3:6, 2: 17 × 6, 3: 21 × 7, 4: 29 × 8, 5: 41 × 13, 6: 30 × 15, $K_s$ by (56), s = 1; : : : ; 6, dipole matrix elements M in e × nm.

| lat.extens. | 1 | 2 | 3 | 4 | 5 | 6 |
|---|---|---|---|---|---|---|
| $M_{0;GS}$ | 0.5 | 0.434 | 0.389 | 0.322 | 0.2246 | 0.241 |
| $f_{GSH}$ | 5.735 | 5.136 | 5.078 | 5.044 | 4.94 | 5.015 |
| $M_{0;ES}$ | 1.63 | 2.34 | 2.625 | 2.88 | 3.09 | 3.1 |
| $f_{ESH}$ | 0.104 | 0.16 | 0.124 | 0.0967 | 0.0738 | 0.0766 |
| $f_{GSL}$ | 5.37 | 4.84 | 4.78 | 4.74 | 4.68 | 4.73 |
| $\Gamma_{GS;\,r}$ (4K) | 2.01 | 2.07 | 2.018 | 2.44 | 4.21 | 3.54 |
| $\Gamma_{GS;\,r}$ (273K) | 4.74 | 4.8 | 4.748 | 5.17 | 6.94 | 6.27 |
| $\Gamma_{ES;\,r}$ (4K) | 2.49 | 4.98 | 6.69 | 10.49 | 32.91 | 24.63 |
| $\Gamma_{GS;\,nr}$ (4K) | 0.016 | 0.0165 | 0.0161 | 0.02 | 0.04 | 0.028 |
| $\Gamma_{ES;\,nr}$ (4K) | 0.02 | 0.04 | 0.053 | 0.1 | 0.184 | 0.14 |
| $K_{sH}$ | 0.19 | 0.9 | 1.11 | 1.53 | 2.82 | 2.41 |

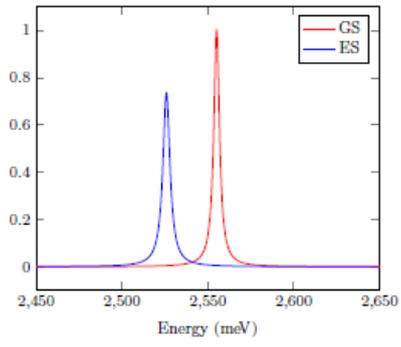

FIG. 1. Normalized PL emission for lateral extension 8.1× 3.6 at 4 K

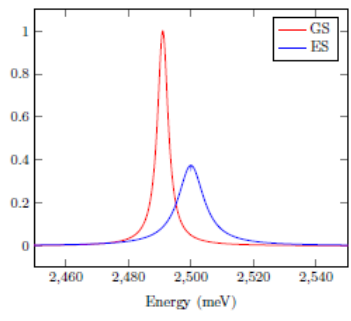

FIG. 2. The same as **Figure** 1, for lateral extension 17× 6

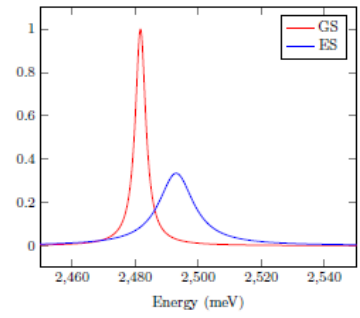

FIG. 3. The same as **Figure** 1, for lateral extension 21× 7

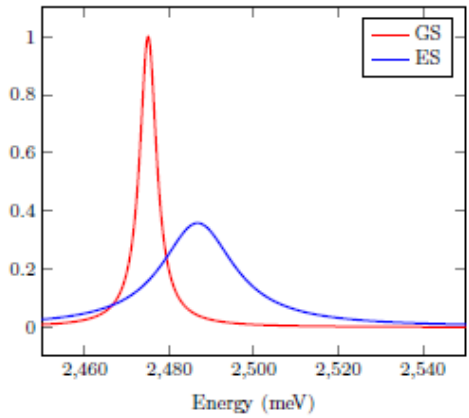

FIG. 4. Lateral extension 29× 8

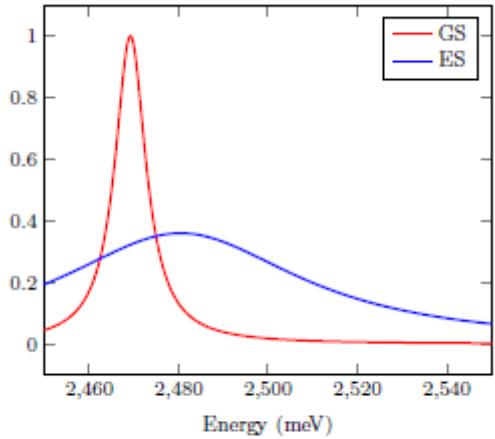

FIG. 5. Lateral extension 41×13

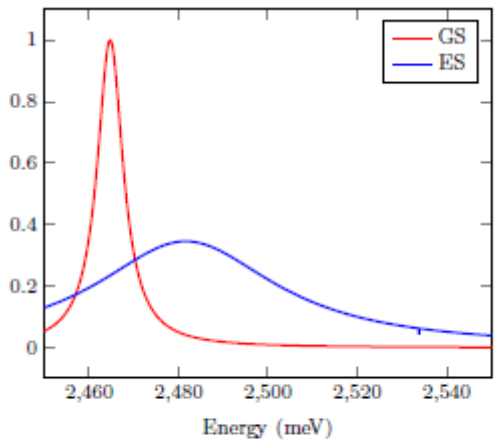

FIG. 6. Lateral extension 30×15

In Ref. [12] absorption spectra at room temperature were measured for two 2 sets of CdSe NPLs, assorted according either to their thickness ('Thickness'), or lateral area ('Lateral Area')

1. Fig. S1 a, suppl.,[12] 'Thickness'
   - 3ML 0:9 ×56 × 41 nm, 'blue',
   - 4ML 1:2 × 17 × 15 nm,'green',
   - 5 ML 1:5×30 × 11 nm, 'red'.

2. Fig. S1 b, suppl.,[12] 'Lateral Area'
   - 4ML 1:2 × 17×15 nm, 'plum',
   - 4 ML 1:2 × 30 × 11 nm, 'olive',
   - 4 ML 1:2 × 56 × 41 nm, 'blue'.

The calculated spectra are shown in Figures 12 and 13. We obtained a fairly good agreement with experimental spectra. The dominating resonances are due to 1SH and 1SL excitons, and are slightly blue-shifted compared with the observed resonance energies. The differences can be explained in the following way. The above given dimensions were described as 'representative'. It means that in experiments different dimensions could be used, which can explain the differences between the theoretical and experimentally measured exciton resonance energies. Moreover, the effective electron and hole masses used through this section were calculated for NPL thicknesses 1, 1.33, and 1.67 nm, whereas in Ref. [12] the thicknesses 0.9, 1.2 and 1.5 nm were used. This also can be a reason for the mentioned differences.

.

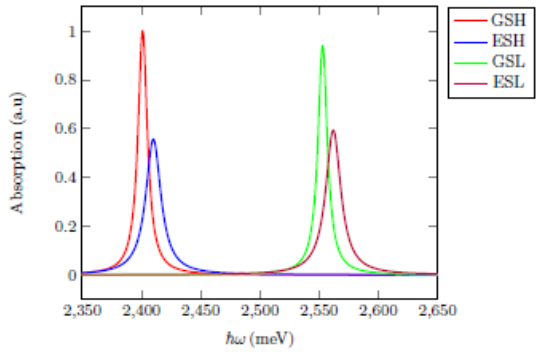

FIG. 7. Normalized absorption, lateral extension $17 \times 6$

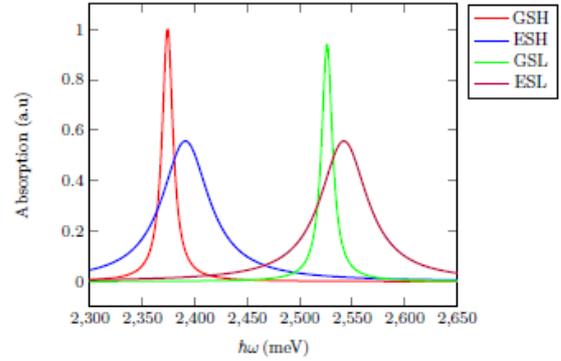

FIG. 10. Normalized absorption, $30 \times 15$

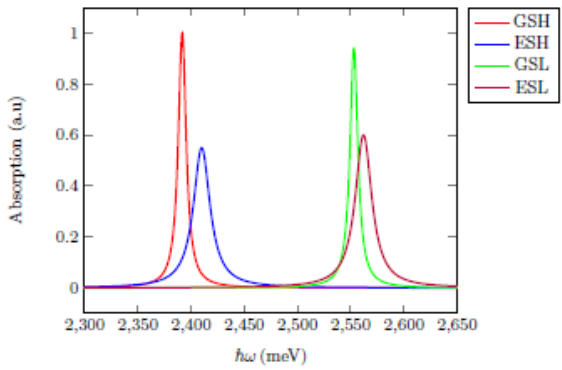

FIG. 8. Normalized absorption, $21 \times 7$

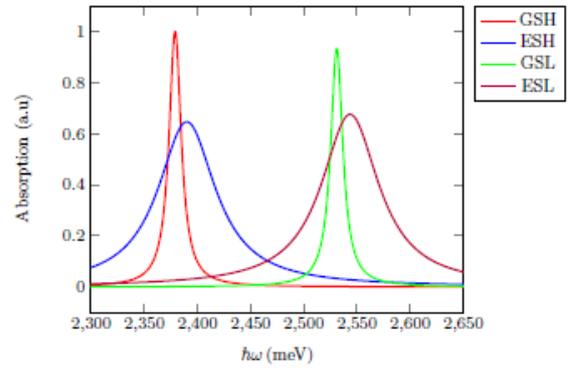

FIG. 11. Normalized absorption, $41 \times 13$

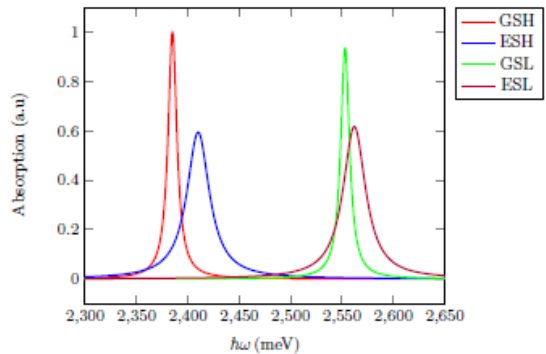

FIG. 9. Normalized absorption, $29 \times 8$

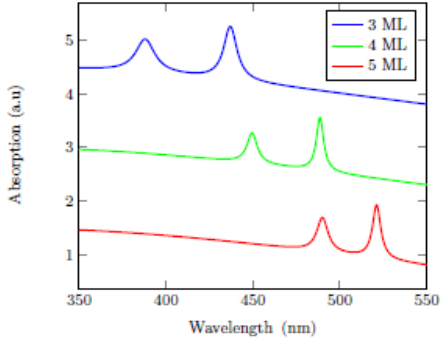

FIG. 12. Normalized absorption for the case 'Thickness'.

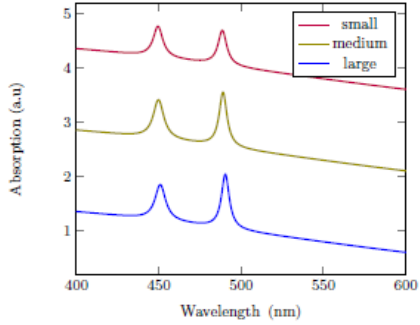

FIG. 13. Normalized absorption for the case 'Lateral area'.

**TABLE 6**. Sizes and exciton states energies, transition matrix elements M, oscillator strengths, and damping parameters, for disks analyzed by Brumberg *et al.* [12], lengths in nm, wave length in nm, matrix elements M in e × nm, energies in meV, the energy gap at room temperature 1750 meV, damping parameter Γ from Eq. (53) notation: 1: 56 × 41; 3ML, 2: 17 × 15; 4ML 3: 30 × 11; 5ML, 4: 17 × 15; 4ML, 5: 30 ×11; 4ML, 6: 56 × 41; 4ML.

| Lat. extens. | 1 | 2 | 3 | 4 | 5 | 6 |
|---|---|---|---|---|---|---|
| $a^*_{e\parallel}$ | 1 | 1.26 | 1.553 | 1.26 | 1.26 | 1.26 |
| $r_{eff}$ | 27 | 9 | 10.25 | 9 | 10.25 | 27 |
| $\rho_{eff}$ | 27 | 7.15 | 6.6 | 7.15 | 8.134 | 21.455 |
| $E_{res}(1SH)$ | 2843 | 2540 | 2381 | 2540 | 2537.7 | 2531 |
| $\lambda$ | 437 | 488.6 | 521.2 | 488.6 | 489 | 490.4 |
| $E_{res}(1PH)$ | 2870 | 2558.48 | 2392.5 | | | |
| $\lambda$ | 432 | 484.6 | 518.7 | | | |
| $E_{res}(1SL)$ | 3201 | 2761 | 2533 | 2761 | 2758.7 | 2752 |
| $\lambda$ | 388 | 449.5 | 490 | 449.5 | 449.8 | 450.9 |
| $M_{0SH}$ | 0.625 | 0.22 | 0.19 | 0.22 | 0.19 | 0.625 |
| $f_{SH}$ | 4.16 | 4.77 | 5.47 | 4.77 | 4.6 | 4.96 |
| $f_{SL}$ | 3.72 | 3.75 | 4.77 | 3.75 | 4.44 | 4.41 |
| $\Gamma_{SH,r}$ | 4.63 | 2.53 | 2.86 | 2.53 | 2.86 | 2.86 |

## 5. Time dependence. Short pulse excitation

Here we refer to the experiments by Achtstein *et al.* [9], where the transient PL decay and evolution of the ES and GS emission with time were examined. The NPLs were irradiated by a short-pulse signal, with the pulse duration of 2 ps. We assume, that the pulse has a Gaussian shape

$$F(t) = F_{max}(1/\tau_p\sqrt{2\pi})\exp(-t^2/2\tau_p^2), \qquad (56)$$

where $\tau_p$ is the pulse temporal duration. Inserting the above pulse shape into Equation (4), we separate the excitonic amplitudes Y into two parts, slow and rapid (see also Ref. [15]). The processes of the excitonic creation are rapid processes, on the femto-second scale. The decline process is proportional to the exciton life time, which is of the order of a few ps, i.e. is a slow process. Moreover, one has two exciton life times, related to radiative (r), and non-radiative (nr) declines. For the slow counterpart of the excitonic amplitudes we obtain two equations

$$dY_{slow,r}/dt - (dY_{slow,r}/dt)_{irrev} = F(t), \qquad (57)$$

$$dY_{slow,nr}/dt - (dY_{slow,nr}/dt)_{irrev} = F(t), \qquad (58)$$

where

$$(dY_{slow,r,nr}/dt)_{irrev} = (-1/\tau_{r,nr})\, Y_{slow,r,nr}\,. \qquad (59)$$

The equations can be solved by means of Green's function, satisfying the equation

$$dG(t,u)/dt + (1/\tau)G(t,u) = -\delta(t-u), \qquad (60)$$

with the solution

$$G(t,u) = (\tau/2)\exp[-(1/\tau)|t-u|]. \qquad (61)$$

Using the function G, we obtain the solutions of Eqs. (57,58) in the form

$$Y_{slow}(t) = \int F(u)\, G(t,u)\, du. \qquad (62)$$

Using the pulse shape (56) we calculate the amplitudes $Y_{slow}$. In the lowest approxi-mation one obtains

$$Y_{slow} = F_{max}\,(\tau/2)\exp(-|t|/\tau). \qquad (63)$$

Making use of Eq. (63) for radiative- and non-radiative decays, we calculate the susceptibility and the absorption coe-fficient, where the time dependent terms have the form

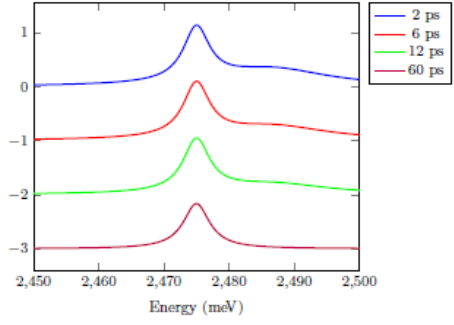

$$\alpha_{GS} \sim f_{GS} \, M^2_{0,GS} / \, \Gamma_{GS} \, \exp(-|t| / \tau_{GS,r}) + B_{GS} \, \exp(-|t| / \tau_{GS,nr}), \quad (64)$$

with quite analogous expression for the excited state. The same equations, with appropriate parameters, hold for the H and L excitons. The decline times follow from the relation

$$\tau_{r,nr} = \hbar / \Gamma_{r,nr} \ . \quad (65)$$

The quantities $\Gamma$ are given in Tables 5 and 6. Below we give, as example, the formula for emission of the NPL with dimensions $29 \times 8 \ \text{nm}^2$

$$\alpha = 6[(2475 - E_{in})^2 + 6]^{-1}[e^{-3.67 \, |t|} + e^{-0.003 \, |t|}] + 40[(2486.7 - E_{in})^2 + 110]^{-1}[e^{-10.94 \, |t|}$$
$$+ e^{-0.05 \, |t|}]. \quad (66)$$

It can be seen, that the long-time contribu-tions are dominant. The excited state declines faster than the ground state. For times t>60 ps only the contribution of the ground state remains. The emission shape for 4 values of time is illustrated in Fig. 14.

Making use of the expression (64), we have calculated the ES and GS transients for different NPL sizes and the pulse shape (56). The results are presented in Figures 15 − 17. We have used the logarithmic scale. The changes due to varying NPL sizes can be noticed. The decline rate is increasing with the lateral area.

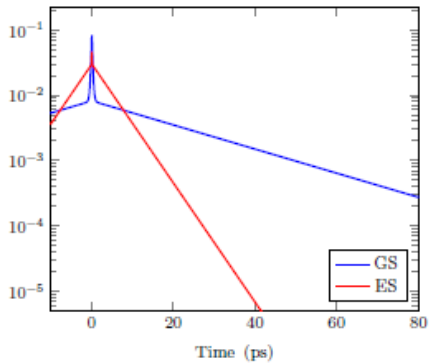

The discussion of the optical functions time dependence can be enlarged by including the temperature dependence. As is known, the general tendency is the increase of the damping parameters with temperature (see, for example, [23]). Since the radiative decline rate increases at least linearly with temperature, we assumed the formula

$$\Gamma_{GS;r} = \Gamma_{GS;r0} + 10^{-2} \, T \, , \qquad (67)$$

with analogous expression for the excited state, T is the temperature . The dependence of the non-radiative decay rates on temperature is, to our best knowledge, not known. We assumed the following fit

$$\Gamma_{GS;nr} = \Gamma_{GS;nr0} + \rho_{eff} \times 10^{-4} \, T^2/(T+235), \qquad (68)$$

where $\Gamma_{GS;nr0}$ is the value at 0 K. The same fit is assumed for the excited state. The effects of temperature are overlapping with the effects of varying NPL sizes. This is illustrated in Figures 18 and 19. The separation of radiative- and non-radiative declines is presented in Figures 20-21.

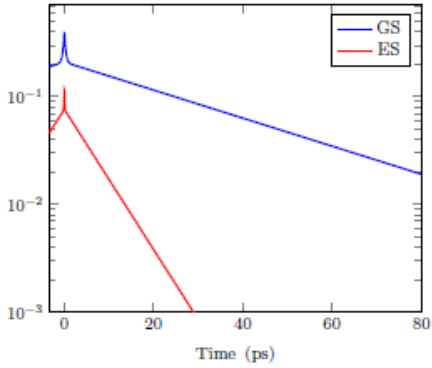

FIG. 16. Evolution of GS emission with time for 2 ps temporal bins platelet size $29 \times 8 \, nm^2$
.

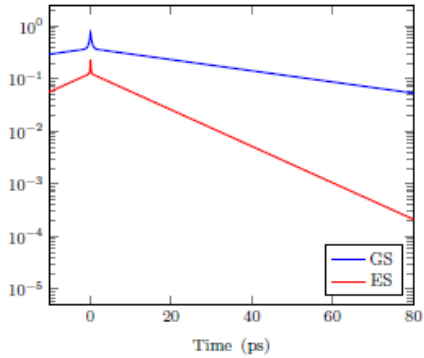

FIG. 17. Evolution of the GS emission with time for 2 ps temporal bins. platelet size $21 \times 7 \, nm^2$

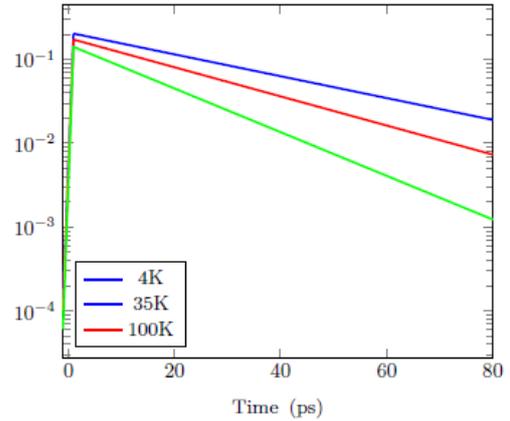

FIG. 18. Evolution of the GS emission with time for 2 ps temporal bins at 4, 35, and 100 K , platelet size $29 \times 8 \, nm^2$

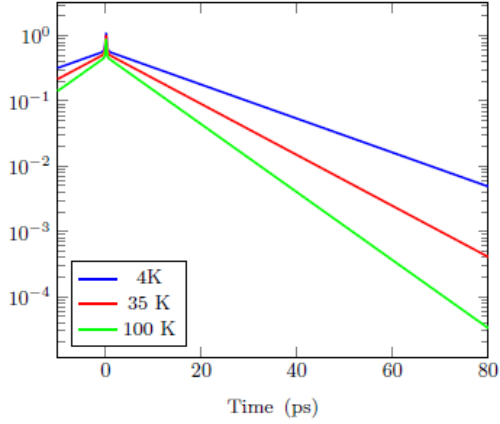

FIG. 19. Evolution of the GS emission with time for 2 ps temporal bins at 4, 35, and 100 K, platelet size $41 \times 13\,\mathrm{nm}^2$

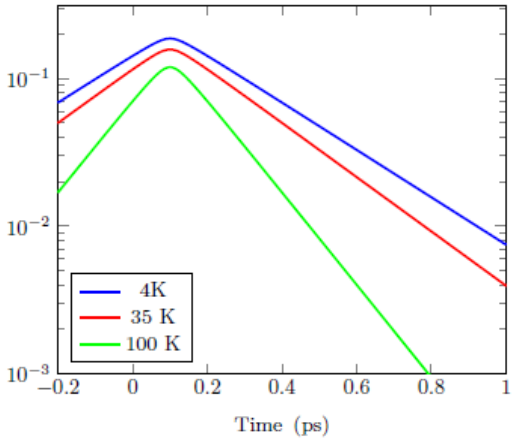

FIG. 20. Evolution of GS emission with time for 2 ps temporal bins, platelet size of $29 \times 8\,\mathrm{nm}^2$

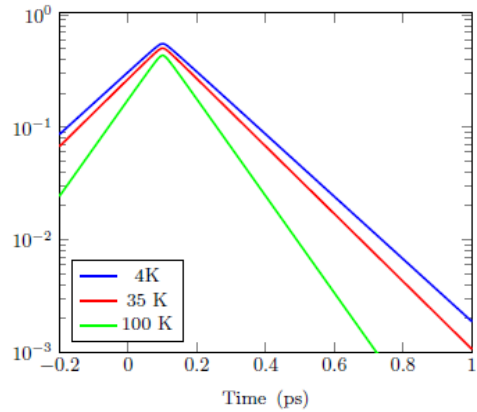

FIG. 21. Evolution of GS emission with time for 2 ps temporal bins platelet size $41 \times 13\,\mathrm{nm}^2$.

## Conclusions

In this paper we have discussed some remarkable optical properties of CdSe monolayers systems. Atomically thin CdSe NPLs have unique physical properties which could be valuable for a broad range of applications [2,24]. Strong light-matter interaction and atomically thin volume are advantages for 2D semicon-ductors which make them easy tunable, as the optical properties can be controlled using multiple modulation methods. The remarkable thinness of these materials also provides unique opportunities for engineering the excitonic properties. For example, changing the dielectric environment of NPLs significantly reduces the exciton binding energies and the free-particle band gap. With the help of RDMA, using dielectric potential resulting from the dielectric confinement, we have derived analytical expressions for the binding energy and absorption for NPLs depending on the monolayers number and lateral area. We also discussed the temperature and time dependence of the spectra, accounting for the radiative- and non-radiative decline rates. Our results have been thoroughly discussed and compared with the available experimental data showing a fairly good agreement. This approach and results may open up a variety of possibilities to manipulate excitonic states on the nanometer scale in 2D materials in the future.

## Appendix: Eigenfunctions and eigenvalues for nanoplatelets with lateral confinement

Below we show how to compute the eigenfunctions and eigenvalues of the Schrodinger equation in the case of one particle mowing in a nanoplatelet, modelled as a disk of radius R, subject to no-escape boundary conditions, and interacting via Coulomb potential with a particle located in the disk center. The quantum mechanical description is given in Eq. (15), and the solutions are given in Eq, (16) (the case of negative energy), and (18) (the case of positive energy). With the purpose of exemplification we consider the cases of 1,2S and 1,2P states. The eigenfunctions include an unknown parameter $\eta$, which then defines the eigenenergy. Thus the first step in calculations consists of the determining $\eta$. Using the definition of the confluent hipergeometric function

$$M(a;b;z)=1+(a/b)z/1!+a(a+1)/[b(b+1)]z^2/2! + \ldots\ldots \qquad (A1)$$

we obtain for the 1S state (j=m=0) the eigenfunction

$$\psi_{00}(\rho) = C_{00}\exp(-\rho/\eta)\ [1+(1/2-\eta)2\rho/\eta + (1/4)\,(1/2-\eta)\,(3/2-\eta)(2\rho/\eta)^2+\ldots], \qquad (A2)$$

where $C_{00}$ is a normalization constant. Retaining the contributions at most quadratic in $\eta$ we have

$$\psi_{00}(\rho) = C_{00}\exp(-\rho/\eta)\ [(1-\rho)^2+(1-2\rho)\,\rho/\eta+ (3/4)\,\rho^2/\eta^2]. \qquad (A3)$$

Substituting $\rho = \rho_{eff}$, we obtain from the boundary condition $\psi_{00}(\rho_{eff}) = 0$ the equation, which determines $\eta$

$$(3/4)\,(\rho_{eff}/\eta)^2 - (2\rho_{eff}-1)\,(\rho_{eff}/\eta) +\ (\rho_{eff}-1)^2 = 0, \qquad (A4)$$

which becomes a quadratic equation for $t = (\rho_{eff}/\eta)$

$$(3/4)\,t^2- (2\rho_{eff}-1)\,t + (\rho_{eff}-1)^2 = 0, \qquad (A5)$$

having the discriminant

$$\Delta =\ \rho_{eff}^2 + 2\,\rho_{eff}-2. \qquad (A6)$$

For $\Delta > 0$ Eq. (A5) has two real and positive roots $t_1$, $t_2$. The condition $\Delta =0$ defines the critical radius $\rho_{cr,00}$ for the 1S states., $\rho_{cr,00} = 0.73$. It means, that for $\rho \geq \rho_{cr,00}$ the 1S states will have negative eigenvalues for the energy. The larger root for t and given $\rho_{eff}$ corresponds to the 1S state, the smaller value gives the energy of 2S state. With the known values of t we obtain the eigenvalues from equations

$$E_{jm} = - (1/\ \eta j_m{}^2)\,R^*{}_{e\parallel}\ \ , \qquad (A7)$$

for the case of negative energies, and

$$E_{jm} = - (1/ \eta j_m^2) \, R^*_{e\parallel} \quad , \tag{A8}$$

for the positive eigenvalues.  An analogous operation can be performed for the 1P (j=0, m=1) state. Leaving only the radial term, one obtains

$$\psi_{01}(\rho) = C_{00} \exp(-\rho/\eta) \, (2\rho/\eta) \, [1 - 2\rho/3 \, + (1/6)\rho^2 - (\rho/\eta) \, (2\rho/3 - 1) \tag{A9}$$

As in the case of 1S exciton, applying the boundary condition we obtain the equation

$$(5/8) \, t^2 - (2\rho_{eff}/3 - 1) \, t + [1 - (2/3) \, \rho_{eff} + (1/6) \, \rho_{eff}^2] \, = \, 0, \tag{A10}$$

## With the same definition of t. Using the discriminant

$$\Delta = (1/36) \, \rho_{eff}^2 + (1/3) \, \rho_{eff} - 3/2 \quad , \tag{A11}$$

we obtain two roots

$$t_{1,2} = 0.8 [2\rho_{eff}/3 - 1 \pm \sqrt{\Delta}]. \tag{A12}$$

The equation $\Delta = 0$ gives the critical radius for the 1P state $\rho_{cr,01} = 3.5$.

When r < $\rho_{cr,01}$, we use the equation

$$\mathrm{Re}\,\psi_{01}(\rho) \, = (2\rho/\eta) \, [(1 - 2r/3) \sin(\rho/\eta) - (\rho/\eta)\cos(\rho/\eta)] [ \, (1/6)\rho^2 - (5/8) \, (\rho/\eta)^2 \, ]. \tag{A13}$$

Substituting

$$\rho/\eta = tx, \qquad x = \rho/\rho_{eff}, \qquad t = \rho_{eff}/\eta \quad ,$$

we obtain the real part of the wave function $\psi_{01}(\rho)$ in the form

$$\psi_{01}(x) = C_{01} \, 2 \, t \, x \, \{ \, [(1 - 2 \, \rho_{eff}/3) \sin(t \, x) - t \, x \cos(t \, x) \, ] \, [(1/6)\rho_{eff}^2 - (5/8) \, t^2 x^2 \, ] \},$$
$$\tag{A14}$$

giving equation

$$[(1 - 2\rho_{eff}/3) \sin(t) - t \cos(t)][(1/6)\rho_{eff}^2 - (5/8) \, t^2] = 0, \tag{A15}$$

for the quantity t. The formulas (A14,15) were applied to the case of 8.1×3.6 NPL, where the effective radius is smaller than the critical radius for 1P excitons.